\documentclass[global,twocolumn]{svjour}
\usepackage[utf8]{inputenc}
\usepackage{amsmath,amsfonts,amssymb,bm}
\usepackage{graphicx}
\usepackage{xcolor}
\usepackage[english]{babel}
\usepackage{subfigure}
\title{A grand-canonical approach to the disordered Bose gas}
\author{Christopher Gaul\inst{1} \and Cord A. M\"uller\inst{2,3}}
\institute{Max Planck Institute for the Physics of Complex Systems, 01187 Dresden, Germany
 \and
 Fachbereich Physik, Universit\"at Konstanz, Germany
 \and
 Centre for Quantum Technologies, National University of Singapore, 117543 Singapore}
\usepackage{enumitem}
\setitemize{topsep=2pt, itemsep=0pt, parsep=2pt, leftmargin=1.5em}
\setenumerate{topsep=2pt, itemsep=0pt, parsep=2pt, leftmargin=1.5em}

\usepackage[sort&compress,numbers]{natbib}
\usepackage[colorlinks]{hyperref}
\usepackage{doi}
\newcommand \enquote  [1]{\emph{#1}}%
\usepackage{dsfont}

\newcommand{\avg}[1]{\overline{#1}}
\newcommand{\mv}[1]{\left\langle #1\right\rangle}
\newcommand{\Hgc}{\hat{H}_\text{gc}}
\newcommand{\vc}[1]{\bm{#1}}
\renewcommand{\r}{{\vc r}}
\renewcommand{\k}{{\vc k}}
\newcommand{\q}{{\vc q}}
\newcommand{\p}{{\vc p}}

\newcommand{\ah}{{\hat a}}
\newcommand{\bh}{{\hat \gamma}}  
\newcommand{\rmd}{{\rm d}}
\newcommand{\rmi}{{\rm i}}
\newcommand{\gh}[1]{\hat\gamma_{#1}}		
\newcommand{\ghd}[1]{\hat\gamma^\dagger_{#1}}	%
\newcommand{\cvect}[2]{\left(\begin{array}{c}#1\\#2\end{array}\right)}         

\newcommand{\Vtil}{\tilde{U}}	

\newcommand{\kB}{{k_{\rm B}}}

\newcommand{\epn}[1]{\varepsilon^0_{#1}}	
\newcommand{\ept}[1]{\tilde{\varepsilon}_{#1}}	%
\newcommand{\ep}[1]{\varepsilon_{#1}}		
\newcommand{\epB}[1]{\varepsilon_{#1}^\text{B}}
\newcommand{\nc}{n_{\rm c}}
\newcommand{\Nc}{N_{\mathrm c}}
\newcommand{\Nczero}{N_{\mathrm c}^{(0)}}
\newcommand{\nB}[1]{\nu_{#1}}			
\DeclareMathOperator{\tr}{tr}
\newcommand{\rmT}{\mathrm{T}}

\date{}

\begin{document}
\maketitle
\begin{abstract}
 We study the problem of disordered interacting bosons within grand-canonical thermodynamics and Bogoliubov theory.
 We compute the fractions of condensed and non-condensed particles and
 corrections to the compressibility and the speed of sound due to interaction and disorder. There are two small parameters, the disorder strength compared to the chemical potential and the dilute-gas parameter.
\end{abstract}

\section{Introduction: grand canonical formalism}

We approach the weakly interacting Bose gas with 
the grand-canonical Hamiltonian \cite{Dalfovo1999,Pethick2002,Pitaevskii2003}
\begin{align}\label{eqManyParticleHamiltonian}
 \Hgc = \int &{\rm d}^3 r\,
\hat \Psi^\dagger \left[\frac{-\hbar^2}{2m} \nabla^2 + U(\r) - \mu + \frac{g_0}{2}\hat \Psi^\dagger \hat\Psi
 \right] \hat \Psi. 
\end{align}
The annihilation (creation) operators $\hat \Psi^{(\dagger)} = \hat
\Psi^{(\dagger)}(\r)$ obey bosonic canonical commutator relations,
$\mu$ is the chemical potential, $g_0>0$ the  
repulsive s-wave interaction strength, and $U(\r)$ an external
one-body potential that renders the gas inhomogeneous. As an application, 
we have in mind either a weak lattice or a random
potential. In the latter case, meaningful quantities will involve the
ensemble average $\avg{(\cdot)}$.  
 
In order to describe the thermodynamic properties of the gas, 
one would like to know the ensemble-averaged grand potential (GP)
$\Omega(\beta,\mu)$, where $\beta = 1/k_{\rm B}T$ is the inverse
temperature: 
\begin{equation}\label{eqOmega}
 {-}\beta \Omega  = \avg{\ln \Xi}= \avg{\ln \{ \tr [ \exp(-\beta \Hgc)
   ] \}}. 
\end{equation}
$\Xi(\beta,\mu)$ is known as the grand partition function. Other than
on $\beta$ and $\mu$, the partition function and the Gibbs state $\hat
\rho = \Xi^{-1}\exp\{-\beta \Hgc\}$ depend
also on all the parameters appearing in the Hamiltonian
\eqref{eqManyParticleHamiltonian}, such as the detailed configuration
of the external potential
$U(\r)$. The grand potential, on the other hand, only contains
those properties that are relevant after the ensemble average. 
The advantage of
this approach is that one obtains relevant physical quantities
directly by
differentiating the GP. 

In particular, the average particle number is
$N(\beta,\mu)=  \tr\{\avg{\hat\rho}\hat N\} =  -\partial
\Omega/\partial \mu$. Often, one prefers to treat intensive quantities
in the thermodynamic limit, such as the
density $N/V=n$. The functional dependence $n(\beta,\mu)$ is known
as the \emph{equation of state}.
By further differentiation, one has access to thermodynamic response
functions, such as the inverse compressibility 
$\kappa^{-1} = n^2 \partial{\mu}/\partial{n}$. 

Due to the interplay of interactions and external potential in
\eqref{eqManyParticleHamiltonian}, though, it is in general impossible to
compute the partition function, let alone the GP,  in closed form without further approximations. Here, we
are interested in the thermodynamics of the Bose-condensed  phase. Therefore, we resort to Bogoliubov's
prescription  $\hat\Psi(\r) = \Phi(\r) + \delta \hat \Psi(\r)$, where
a macroscopically occupied condensate mode $\Phi(\r)$ is separated
from the quantum fluctuations $\delta\hat\Psi(\r)$. 
The condensate plays a
role analogous to the classical trajectory in Feynman's path-integral
formulation of quantum mechanics: on the mean-field level, $\Phi(\r)$ is an extremum, actually a minimum, of the 
functional \eqref{eqManyParticleHamiltonian}
inside  the trace \eqref{eqOmega}. The minimization condition is known as the
Gross-Pitaevskii or non-linear Schr\"odinger equation.  
Including contributions from the quantum fluctuations, one is
later led to minimize more generally the grand-canonical energy, and thus finds
beyond-mean-field corrections to the equation of state.

In this article, we explore the consequences brought about by 
quantum fluctuations in the presence of an external potential
$U(\r)$. These corrections to the `classical' mean-field solution can
be computed by a quadratic expansion of the Hamiltonian \eqref{eqManyParticleHamiltonian} and
subsequent Gaussian integration for the grand potential
\eqref{eqOmega}. 
We take advantage of the effective impurity-scattering Hamiltonian derived in
\cite{Gaul2011_bogoliubov_long} to take into account the external
potential's effect on the condensate (``condensate deformation'')
as well as on the fluctuations.  
Specifically, we derive
 beyond-mean-field corrections to the particle density
(``condensate depletion'') in a disordered Bose fluid at finite
temperature, thus complementing 
the zero-tempera\-ture results of
Ref.~\cite{Muller2012_momdis}. Furthermore, 
we compare some of our findings on the mean-field level with recent measurements \cite{Krinner2013}.  

Before tackling the general, inhomogeneous case in Sec.~\ref{secInHom}, we find it 
instructive to first introduce our readers to the subtleties of grand-canonical
Bogoliubov theory in the homogeneous case, treated in the following
Sec.~\ref{secHom}.  Notably, it is shown how to recover the celebrated
beyond-mean-field corrections to the equation of state first derived
by Lee, Huang, and Yang \cite{Lee1957}.   

\section{Homogeneous case}\label{secHom}

In the homogeneous case  $U(\r)=0$, 
condensation occurs in the $\k=0$ mode \cite{Penrose1956}. Therefore, we only need to
determine the population $\Nc$ of that mode, but not its shape. 
The Bogoliubov approximation consists in replacing the condensate
field operator with a c-number, $\ah_0 = \Nc^{1/2}$, and treating all other $\k$-space
modes as quantum fluctuations. In the following, we will first
establish the effective Hamiltonian, then determine the GP, and
analyze in detail the ground-state density. We close this section with
a discussion of the condensate fraction at zero and finite
temperature. 

\subsection{Hamiltonian} 
 
Expanding the Hamiltonian $\Hgc= H_0 
+ \hat H_2+\dots$ to second order in the fluctuations ($\hat
H_1$ vanishes by momentum conservation), one finds 
\begin{equation}\label{Ec} 
 H_0=  \Nc \left[
  \frac{g_0 \nc}{2} - \mu\right]
  \end{equation}  
  for the mean-field energy, where 
$\nc=|\Phi_\text{c}|^2= \Nc/V$ is the condensate density. If one
 minimizes the mean-field energy alone, $\partial H_0/\partial \nc=0$,
 one finds $g_0\nc=\mu$, and recovers canonical
 Bogoliubov theory \cite{Pethick2002,Pitaevskii2003}. Here, we postpone
 the minimization until the complete GP is known, in order to obtain beyond-mean-field corrections. 

The fluctuations are described by the Hamiltonian 
\begin{align}
\hat H_2 & =  {\sum_\k}^\prime \bigl[ 
   (\epn{\k} + 2g_0\nc -\mu) \ah_\k^\dagger\ah_\k  
   + \frac{g_0 \nc }{2} (\ah_\k^\dagger\ah_{-\k}^\dagger + \text{h.c.}
   )
  \bigr]. \label{eqHomHG_a}
\end{align}
The primed sum indicates that $\k=0$ is 
omitted, and $\epn{\k}$ is the single-particle
dispersion.\footnote{
We note $\epn{\k}$ with a vector index to cover cases where the dispersion is
  anisotropic, e.g., in a tight-binding lattice
  \cite{Gaul2013_bogolattice}. For concrete examples within this paper, we 
  consider only the free-space case, where $\epn{\k}=\hbar^2k^2/2m$ is isotropic. 
In all cases, we assume parity invariance, $\epn{\k}=\epn{-\k}$.}
In order to avoid a UV divergence of the ground-state energy later on, one 
renormalizes the interaction constant in \eqref{Ec} as $g_0 = g + \sum_\k'g^2/2\epn{\k}V$
\cite{Pitaevskii2003}, which adds a c-number term under the sum in
\eqref{eqHomHG_a}. 
The quadratic Hamiltonian $\hat H_2$ becomes  
diagonal after a transformation to the Bogoliubov quasiparticles
$\bh_\k = u_\k \ah_\k + v_\k \ah_{-\k}^\dagger$ and $\bh_\k^\dagger = u_\k \ah_\k^\dagger +
v_\k \ah_{-\k}$, with 
\begin{align}\label{BgCoefficients}
u_\k = \frac{\ep{\k} + \ept{\k}}{2 (\ep{\k}\ept{\k})^{1/2}},\quad 
v_\k = \frac{\ep{\k} - \ept{\k}}{2 (\ep{\k}\ept{\k})^{1/2}},
\end{align}
defined in terms of 
\begin{align} 
\ept{\k} & = \epn{\k} + g \nc - \mu, \label{ept}\\
\ep{\k} & = \sqrt{(\epn{\k} + 2g \nc -\mu)^2 - (g \nc)^2}. \label{ep} 
\end{align} 
All these quantities still depend separately on the
 chemical potential $\mu$ and the condensate population $\nc$.
Only with the choice $g\nc=\mu$, the energy \eqref{ep} 
becomes purely real and gapless, as it should according to 
a theorem by Hugenholtz and Pines \cite{Hugenholtz1959}, and 
turns into the celebrated Bogoliubov dispersion relation,   
\begin{equation} \label{bogospec} 
\epB{\k}=  \sqrt{\epn{\k}(\epn{\k} + 2\mu )}.
\end{equation}
Yet, in order to be able to differentiate with respect
to $\mu$ at fixed $\nc$ (or vice versa), we keep 
both quantities   
and remember to choose $g\nc=\mu$ in all
final expressions relating to the excitations.

Thus, the grand-canonical Hamiltonian takes the form
\begin{align}
 \Hgc = E_0
 + {\sum_\k}^\prime \ep{\k} \bh_\k^\dagger\bh_\k. 
\label{eqHomHG_b}
\end{align}
The first term is the grand-canonical candidate for the ground-state
energy, to which fluctuations contribute with their commutators: 
\begin{equation}  \label{E0}
E_0 = \Nc \left [ \frac{g \nc}{2} - \mu\right] 
  -\frac{1}{2}  {\sum_\k}^\prime \Bigl[ (\epn{\k} + 2g \nc -\mu) -\ep{\k} - \frac{(g \nc)^2}{2\epn{\k}}  \Bigr].
\end{equation} 
At this point, we still have the freedom to choose the condensate
density $\nc$ 
by minimizing the energy \eqref{E0} 
at fixed $\mu$.  Requiring $\partial E_0/\partial\nc|_\mu=0$ results in
\begin{align}
 \nc(\mu) = \frac{\mu}{g} - \frac{5\sqrt{2}}{12\pi^2} \frac{1}{\xi^3},
\label{ncofmu}
\end{align}
where we have introduced the characteristic length $\xi$ via $\mu =
\hbar^2/(2m\xi^2)$. Inserting this result in \eqref{E0}---at mean-field
precision inside the fluctuation sum---yields the
ground-state energy density  
\begin{align}
 \frac{E_0(\mu)}{V} = - \frac{\mu^2}{2g} + \frac{2\sqrt{2}}{15\pi^2}
 \frac{\mu}{\xi^3}. 
\label{E0ofmu}
\end{align}

\subsection{Grand potential} 

The ground-state energy $E_0(\mu)$ thus
determined is a constant in the Hilbert
space of Bogoliubov excitations and 
pulls out of the trace \eqref{eqOmega} for the GP, which evaluates in the thermodynamic limit to  
\begin{align}
\Omega  & = E_0 +  \frac{V}{\beta}  \int \frac{{\rm d}^3
  k}{(2\pi)^3} \ln(1-e^{-\beta \ep{\k}}) . 
\label{eqOmegaHom}
\end{align}
The density derives 
as 
\begin{equation} \label{n_hom_ofmu}
n = 
-\frac{1}{V}\frac{\partial E_0}{\partial\mu} -  \int
\frac{{\rm d}^3 k}{(2\pi)^3}  \nB{\k} \frac{\partial \ep{\k}}{\partial
  \mu}, 
\end{equation} 
where $\nB{\k} = [e^{\beta \ep{\k}}-1]^{-1}$ is the Bose-Einstein
distribution function for the occupation of excitation modes. 
Remember that the second contribution, namely the thermal contribution of fluctuations, should be
differentiated with respect to $\mu$ at fixed $g\nc$, and then
evaluated with $g\nc=\mu$ at the end. 

Alternatively, it is also possible to derive the condensate density
$\nc$ not from the ground-state energy \eqref{E0} (i.e., at zero temperature), but by minimizing
the full GP, eq.~\eqref{eqOmegaHom}, as function of $\mu$ and
arbitrary $T$. Thus, one is able to account for the thermal depletion
of the condensate at fixed $\mu$. The final results (as presented in
the following) come out the same, as described in the Appendix
\ref{appHom}. But for technical reasons that will become apparent in
Sec.~\ref{secInHom} below, we prefer to use the `semicanonical'
prescription, in which $\nc$ is kept independent from $\mu$ when
differentiating, and only substituted later at the required precision.


\subsection{Zero-temperature equation of state} 
The density  $n=-V^{-1}\partial
E_0/\partial\mu$ derived  from \eqref{E0ofmu} thus determines the
zero-temperature equation of state    
\begin{equation} \label{n0_of_mu}
n(T=0,\mu) = \frac{\mu}{g} - \frac{\sqrt{2}}{3\pi^2\xi^3}.   
\end{equation} 
The difference between this total density and the condensate density
\eqref{ncofmu} is the so-called quantum depletion, 
\begin{equation}\label{deltan0} 
\delta n_0 = n-\nc = \frac{\sqrt{2}}{12\pi^2\xi^3}. 
\end{equation} 
The depletion must be small compared to $n$ (and thus $\nc$) in
order for the Bogoliubov ansatz to hold. In this case we can express $\xi =
(8\pi n a)^{-1/2}$ through the s-wave scattering length $a$ and total
density $\nc\approx n$ itself, and recover the 
equivalent canonical expression \cite[eq.\ (4.34)]{Pitaevskii2003}
\begin{equation}
 \mu = g n \Bigl( 1 + \frac{32}{3} \sqrt{n a^3/\pi}\Bigr). \label{eqEqOfState}
\end{equation}
Here, the dilute-gas parameter $\sqrt{n a^3}$ has come into play,
which must be 
small for this correction to be meaningful.

One can further derive the compressibility
\begin{equation} 
\kappa^{-1} = n^2\frac{\partial \mu}{\partial n} =  g n^2 \left(1+16\sqrt{n a^3/\pi}\right). 
\end{equation}  
The corresponding speed of sound, determined by $\kappa^{-1}= n m
c^2$
\cite{Leggett2001,Ueda2010},
\begin{equation} 
c= \sqrt{\frac{g n }{m}} \left(1+8\sqrt{n a^3/\pi}\right),
\end{equation} 
includes the Lee-Huang-Yang correction.%
\footnote{There is a misprint in the original paper \cite{Lee1957}:
In eq.\ (33), the square root of the expression in bracket is missing, i.e., the relative correction to the speed of sound should be $8\sqrt{n a^3/\pi}$, not $16\sqrt{n a^3/\pi}$. Unfortunately, this mistake has been copied in the book by Ueda \cite[eq.\ (2.57)]{Ueda2010}. 
The correct result is given, for example, in \cite[eq.\
(2.23)]{Fetter1971}, \cite[eq.\ (25)]{Ronen2009}, and \cite[eq.\
(1.149)]{GueryOdelin_LesHouches2009}.} 

\subsection{Condensate fraction at zero and finite temperature}\label{secCondensateFraction_clean}
\begin{figure}[bt]
 \includegraphics[height=0.56\linewidth]{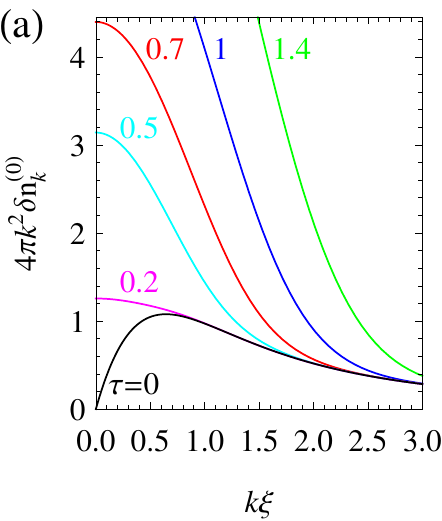}
 \includegraphics[height=0.56\linewidth]{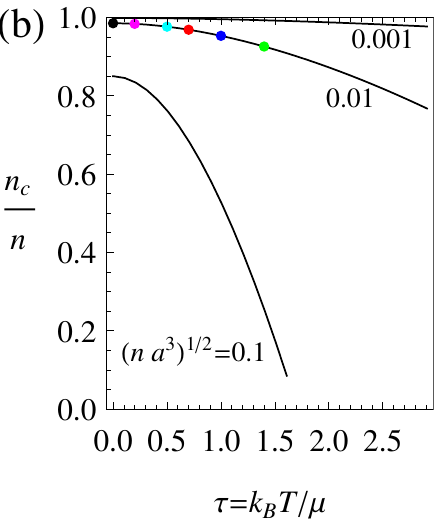}
\caption{Clean condensate fraction and depletion in 3D. (a) Single-particle momentum distribution for reduced  temperatures $\tau = \kB T/\mu = 0, 0.2, 0.5, 0.7, 1.0, 1.4$. (b) Condensate fraction $\nc/n$ [eq.~\eqref{eqParticleNumberTh}] as function of temperature for different values of the dilute-gas parameter $(n a^3)^{1/2}=0.1,0.01,0.001$.
\label{figClean}}
\end{figure}

Often, one is interested in the condensate fraction $\nc/n$ as
function of temperature and fixed total density. 
Equation \eqref{n_hom_ofmu} with the help of eq.\ \eqref{E0} gives the well-known formula \cite[eq.\ (4.42)]{Pitaevskii2003} 
\begin{equation}
\frac{\nc}{n} =1 
 - \frac{2^{3/2}(n a^3)^{1/2}}{\pi^{3/2}} 
     \int{{\rm d}^3 (k\xi)} \left[v_\k^2 +  (u_\k^2+v_\k^2) \nB{\k}\right] , \label{eqParticleNumberTh}
\end{equation} 
Figure \ref{figClean} visualizes this result by showing
(a) the integrand, or single-particle momentum distribution 
$\langle \ah^\dagger_\k \ah_\k\rangle = v_\k^2 + (u_\k^2+v_\k^2) \nB{\k}$, as function of reduced momentum $k\xi$
for different temperatures and (b) the resulting condensate fraction as function of temperature. 
Bogoliubov theory can be expected to give reasonably accurate results when the condensate fraction is large, i.e., for weak interaction and low temperatures.

\section{Inhomogeneous (disordered) case}
\label{secInHom}

The presence of an external potential substantially complicates the
situation, especially if $U(\r)$ is a random function. For a given
realization, the bosons condense into a macroscopically populated
eigenmode of the one-body density matrix \cite{Penrose1956}, whose precise
form is shaped by the interplay of kinetic, interaction, and potential energy. 

In the spirit of Bogoliubov theory, one first needs to find
the deformed condensate amplitude, given as a functional $\Phi(\r)=\Phi[U(\r)]$ and depending
of course also on $\mu$ and $g$. The total occupation number of this mode, 
\begin{equation} \label{Nc} 
\Nc = \int\rmd^3 r |\Phi(\r)|^2 = \sum_\k |\Phi_\k|^2,
\end{equation}
is now larger than the occupation $N_0 = |\Phi_0|^2$ of the coherent
mode $\k=0$ alone \cite{Astrakharchik2011}. The inhomogeneous
components $\Phi_\k = V^{-1/2} \int\rmd^d r e^{-\rmi\k\cdot\r}
\Phi(\r)$ with $\k\neq 0$ describe a  
``deformed condensate'' \cite{Muller2012_momdis} or ``glassy fraction''
\cite{Yukalov2007}. In a second step, one may then describe the quadratic 
fluctuations around this deformed condensate. We are assured to
find a well-defined set of elementary excitations whenever the
external potential is weak enough not to fragment the condensate. 

In this section, we first determine the condensate amplitudes for a
given external potential on the mean-field level, and thus derive disorder corrections
to the mean-field equation of state of the previous section. Our
prediction for the resulting dependence of the compressibility on
disorder strength compares rather well with recent measurements
with ultracold molecules confined to 2D in presence of laser-speckle
disorder \cite{Krinner2013}. 

In a second step, we put the quadratic Hamiltonian of the fluctuations
to use and calculate their contribution to the GP. From there, we
derive disorder corrections to the condensate depletion, recovering
the zero-temperature results of \cite{Muller2012_momdis} and extending
them to finite temperatures.

\subsection{Mean-field equation of state and compressibility}
As before, 
we expand the grand-canonical Hamiltonian 
$\Hgc = H_0      + \hat H_2 + \ldots$
up to second order in the fluctuations.
On the mean-field level, the condensate amplitude minimizes the
Gross-Pitaevskii functional, which reads in momentum representation  
\begin{align}\label{H0disorder}
H_0 &=\sum_{\k\k'}\Phi^*_{\k}
 \bigl[(\epn{\k}-\mu)\delta_{\k \k'} + U_{\k-\k'}\bigr]
 \Phi_{\k'} \nonumber \\
 &\quad+\frac{g}{2V}\sum_{\k\p\k'} \Phi^*_{\k}\Phi^*_{\p-\k}\Phi_{\p-\k'}\Phi_{\k'} 
\end{align}
and thus generalizes \eqref{Ec}.
For a weak external potential, whose smoothed Fourier components \cite{Sanchez-Palencia2006}
\begin{equation}
\Vtil_\k = U_\k/(2\mu+\epn{\k})
\end{equation}
are a set of small numbers,  
one can compute a perturbative solution $\Phi_\k =\Phi_\k^{(0)}+\Phi_\k^{(1)}+\Phi_\k^{(2)}+\dots$ around the homogeneous condensate $ \Phi^{(0)}_\k = \phi_0\delta_{\k 0}$ 
with $\phi_0^2= \Nczero=V\mu/g$  
\cite{GaulPhD2010}:
\begin{align}
 \Phi^{(1)}_\k & = - \phi_0 \Vtil_\k , \label{smoothing01}\\
 \Phi^{(2)}_\k &= - 
 \phi_0 \sum_{\k'} \frac{\mu-\epn{\k-\k'}}{2\mu+\epn{\k}} \Vtil_{\k'}\Vtil_{\k-\k'}. \label{smoothing2}
\end{align}
Using this solution in \eqref{H0disorder}, we find the ground-state GP 
\begin{equation}\label{OmegaDisorderMF}
\frac{\Omega_0}{V}= 
\frac{\avg{H_0}(\mu)}{V} = - \frac{\mu^2}{2g} +\frac{\mu U_0}{g}- 
\frac{\mu}{g}\sum_\k\frac{\avg{|U_\k|^2}}{\epn{\k}+2\mu}. 
\end{equation} 
By virtue of $\avg{\nc}=-V^{-1}\partial \Omega_0/\partial\mu$, or by  
inserting the perturbative solution
\eqref{smoothing01}--\eqref{smoothing2} directly into \eqref{Nc}, the ensemble-averaged equation of state becomes
\begin{equation} 
\label{gnc_of_muU} 
g \avg{\nc}(\mu) = \mu-U_0 + \sum_\k  \frac{\epn{\k}\avg{|U_\k|^2}}{(\epn{\k}+2\mu)^2}. 
\end{equation} 
The first-order effect of the external potential is to shift the
chemical potential by its mean value $\avg{U(\r)}=U_0$. To second
order, at fixed $\mu-U_0$, the external potential draws more
particles into the condensate. 

\begin{figure}
\begin{center}
\includegraphics[width=0.75\linewidth]{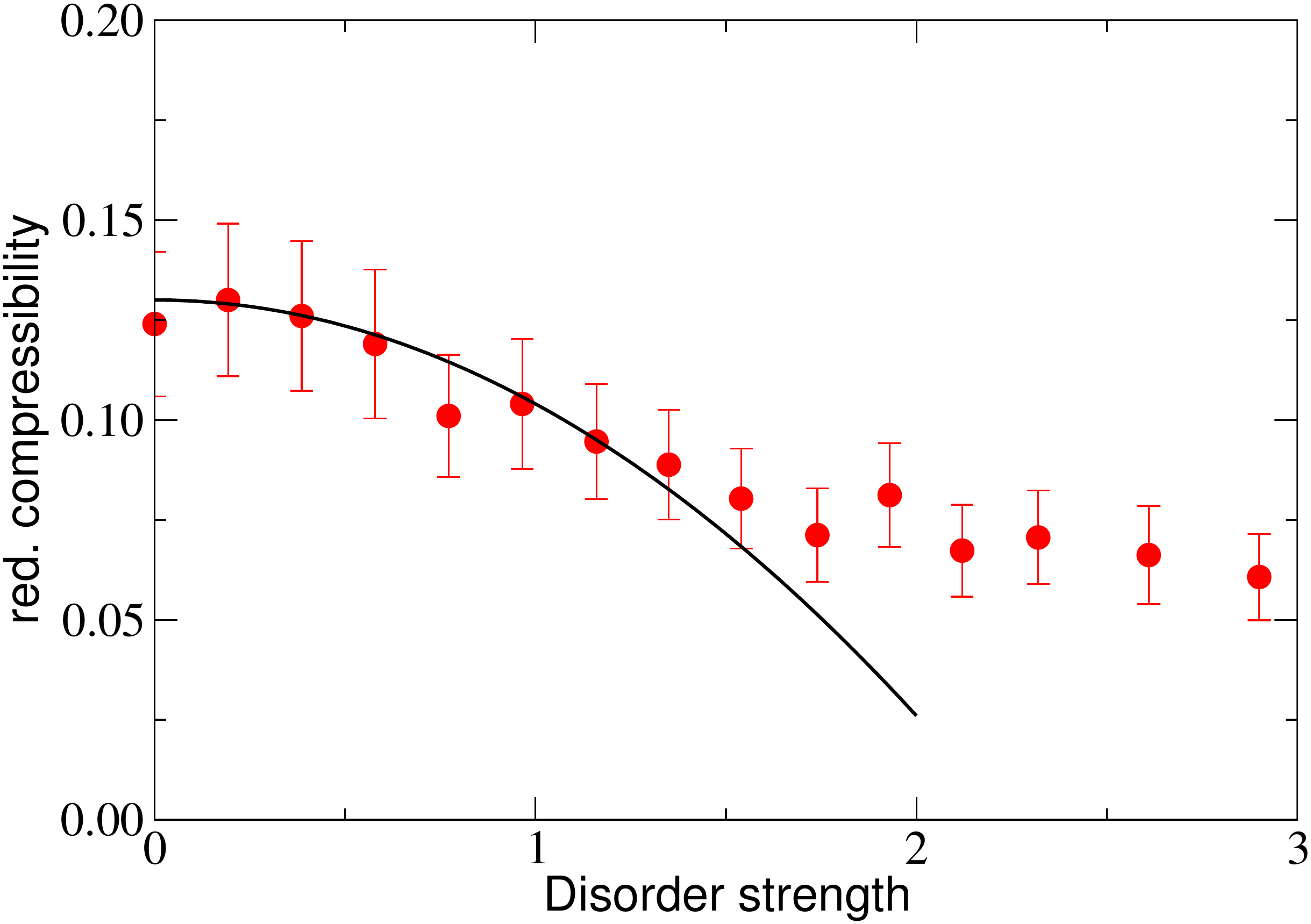}
\caption{Reduced 2D compressibility $\bar\kappa = (\hbar^2/m)\partial
  \avg{n}/\partial\mu$ as function of disorder strength $U/\mu$. The data points
  are measured values from Ref.~\cite{Krinner2013} (courtesy of
  J.-P.~Brantut). The solid line is the result of \eqref{compC} in 2D
  for $\sigma=\xi$ and a Gaussian correlation of the type
  \eqref{gaussian}, scaled to match the disorder-free value for
  $U=0$.}
\label{comprfig}
\end{center} 
\end{figure}

We thus find the mean-field compressibility  
\begin{equation} \label{compU} 
\kappa^{-1}_\text{c} = g n^2 \left(1+4 \sum_\k \frac{\epn{\k}
    \avg{|U_\k|^2}}{(\epn{\k}+2 g n)^3}
  \right). 
\end{equation}  
In all of the preceding expressions appears the pair-correlation
function (with $V=L^d$ in $d$ dimensions) 
\begin{equation}
\avg{|U_\k|^2} = U^2 \frac{\sigma^d}{V} C(\sigma \k),   
\end{equation}
which contains information about the strength of disorder, via the
variance $U^2$ of on-site fluctuations. It also specifies spatial
correlations, via the correlator $C(\sigma\k)$. This function
typically decays over a spatial correlation length $\sigma$, which is of the order of a micron
in experiments involving laser speckle. Then, \eqref{compU} can be
written 
\begin{equation} \label{compC} 
\kappa^{-1}_\text{c} = g n^2 \left(1+4 \frac{U^2}{\mu^2} \int \frac{\rmd^d(\sigma k)}{(2\pi)^d}
  \frac{k^2\xi^2 C(\sigma\k)}{(k^2\xi^2+2)^3}
  \right). 
\end{equation} 
Thus, the compressibility is expected to
decrease quadratically with increasing disorder strength
$U/\mu$ at fixed correlation ratio $\sigma/\xi$.  
The compressibility can be measured with some precision in cold-atom
experiments such as \cite{Krinner2013}. There, a quadratic
decrease of the compressibility is measured for weak disorder, 
in quantitative agreement with \eqref{compC}, when evaluated in
2D  with the correlation length 
$\sigma\approx \xi$ comparable to the healing length, as shown in
Fig.~\ref{comprfig}. Since the 2D molecular BEC in the experiment is
rather strongly interacting, with a depletion of order unity already
without disorder, beyond-mean-field corrections to the homogeneous
compressibility $\kappa^{(0)}_\text{c} = 1/g n^2$ are important. The plotted correction
$\bar{\kappa}^{(0)} (1 - \alpha U^2)$ therefore starts from the
experimentally measured value for $\bar\kappa^{(0)}$. The agreement is
satisfactory for disorder strengths $U/\mu$ not exceeding unity, as
expected for a random potential whose correlation length is of the
order of the healing length itself. 

Comparing the two equations of state considered so far, \eqref{n0_of_mu} and
\eqref{gnc_of_muU},  we see that there are two small parameters: the
dilute-gas parameter $\sqrt{n a^3}$, and the dimensionless disorder
strength $U^2/\mu^2$. Within the scope of this article, we are
interested in their first-order effects. Therefore, we do not consider
cross terms that would come from a higher-order solution of the ground
state energy (the equivalent of \eqref{E0} including contributions from
the fluctuations), which is a somewhat ill-defined quantity in the
presence of disorder anyway.%
\footnote{Notably, it does not seem evident how to implement
  a counterterm that guarantees the convergence of \eqref{E0} in
  presence of disorder \cite{Falco2007}.}  
Also, we do not attempt to derive the condensate amplitude by
minimizing the full GP including the fluctuations (to be described
shortly), and thus forego a direct access to the thermal depletion of
the condensate. But as the homogeneous case showed, we are allowed
to use the `semicanonical' method by keeping $\mu$ and condensate
$\Phi(\r)$ formally independent, and differentiating with respect to
$\mu$ alone, inserting the mean-field solution $\Phi(\r)$ to the
required precision at the end. In the
following, we take into account the 
first-order effect of disorder via a compensating shift in $\mu$, and can assume without loss of
generality that the potential has zero mean, $U_0=0$, such that only
second-order corrections need to be discussed.

\subsection{Quadratic Fluctuation Hamiltonian}
Turning now to the quantum fluctuations, also affectionately called ``bogolons'', we first must ensure that they live in the space orthogonal to the condensate \cite{Fetter1972}.
This constraint can be respected in the density-phase parametrization 
(see also \cite{Mora2003} for a number-conserving approach, and \cite{Lugan2011,GaulSchiefele2014} for the connection between both approaches), where the excitations are given by \cite{Gaul2011_bogoliubov_long,Muller2012_momdis}
\begin{equation}\label{delPsikgamma}
\delta\hat\Psi_\k = \sum_\p \left(u_{\k\p}\gh{\p} -
  v_{\k\p}\ghd{-\p}\right). 
\end{equation}
The transformation matrices
\begin{align}
u_{\k\p} & = \frac{1}{2\phi_0}\left[ a_\p^{-1}\Phi_{\k-\p} +
    a_\p\check\Phi_{\k-\p} \right], \label{ukp}\\
v_{\k\p} & =  \frac{1}{2\phi_0}\left[ a_\p^{-1}\Phi_{\k-\p} -
    a_\p\check\Phi_{\k-\p} \right],   \label{vkp}
\end{align}
contain the Fourier coefficients $\Phi_\k$
of the condensate
amplitude $\Phi(\r) $ and its inverse $\check\Phi(\r) =\nc/\Phi(\r)$, which encode the 
dependence on the external potential $U(\r)$ or rather
its Fourier components $U_\k$, as described in \eqref{smoothing01} and \eqref{smoothing2}. 
In the absence of an external
potential, the condensate $ \Phi^{(0)}_\k =\phi_0\delta_{\k 0}$
renders this transformation diagonal in $\k$, and by choosing $a_\k=(\ept{\k}/\ep{\k})^{1/2}$, 
one recovers the homogeneous transformation \eqref{BgCoefficients}.

Now, we seek the effective quadratic Hamiltonian that describes
these excitations. 
As the condensate mode $\Phi$ minimizes $H_0$, the linear term in the
expansion vanishes and the relevant term is the
second-order fluctuation Hamiltonian 
\begin{equation} \label{H2inhom} 
\hat H_2 = \hat H_2^{(0)} + \hat{U},
\end{equation} 
 where $\hat
H_2^{(0)} = \sum_{\k}\ep{\k}\ghd{\k}\gh{\k}$ formally looks like the free-space
contribution---but please be reminded that the excitations defined via 
\eqref{delPsikgamma} are \emph{not} the
plane-wave modes of the homogeneous case. Furthermore, we recall that  $\mu=g\nc$ has to be
taken in expressions relating to the fluctuations at the end, thus
ensuring a real, gapless excitation spectrum. And finally, we have
discarded the zero-point contribution of commutators, which would
result, as explained above, in a beyond-mean-field modification of the
ground-state energy, which is not investigated here. 
More importantly, the spatial inhomogeneity
leads to the appearance of the scattering potential  
\begin{align}\label{eqInhomHG2}
\hat U &= 
\frac{1}{2} {\sum_{\k,\k'}}'
(\ghd{\k},\gh{-\k})  
\begin{pmatrix} W_{\k\k'}& Y_{\k\k'}\\ Y_{\k\k'}& W_{\k\k'}\end{pmatrix} \cvect{\gh{\k'}}{\ghd{-\k'}}  . 
\end{align}
The impurity-scattering matrices  
\begin{align}
W_{\k\k'} &= \tfrac{1}{4} \left[a_\k a_{\k'} R_{\k\k'} +
  a_\k^{-1}a_{\k'}^{-1} S_{\k\k'}  \right] - \delta_{\k \k'}\ep{\k}, \label{eqW} \displaybreak[0]\\ 
Y_{\k\k'} &= \tfrac{1}{4} \left[a_\k a_{\k'} R_{\k\k'} -
  a_\k^{-1}a_{\k'}^{-1} S_{\k\k'}  \right], \label{eqY} 
\end{align} 
can be traced back to the terms $h_{\k\k'}= (\epn{\k}-\mu)\delta_{\k \k'} + U_{\k-\k'}+2g
{\nc}_{\k-\k'}$ and $g {\nc}_{\k-\k'}$ in the inhomogeneous
generalization of eq.~\eqref{eqHomHG_a}:
\begin{align}
S_{\k\k'} &= \frac{2}{\phi_0^2}{\sum_{\p \q}} 
  \Phi_{\k-\p}\bigl[
h_{\p\q} - g {\nc}_{\p-\q}
\bigr] \Phi_{\q-\k'} , \label{eqS} \displaybreak[0]
\\
R_{\k\k'} &= \frac{2}{\phi_0^2} {\sum_{\p \q}}
  \check\Phi_{\k-\p}\bigl[
h_{\p\q} + g {\nc}_{\p-\q}
\bigr] \check\Phi_{\q-\k'} . \label{eqR}
\end{align}
In contrast to the otherwise equivalent formulas in Refs.\
\cite{Gaul2011_bogoliubov_long,Muller2012_momdis,Gaul2013_bogolattice},
we keep the chemical potential $\mu$ and the condensate mode
[$\Phi(\r)=\sqrt{\nc(\r)}=\nc/\check\Phi(\r)$] separately here, 
in order to be able to take partial derivatives with respect to $\mu$
only. 
By virtue of the perturbative expansion
\eqref{smoothing01} and \eqref{smoothing2}, the effective scattering
potential $\hat U = \sum_{n=1}^\infty \hat U^{(n)}$ can be expanded in
powers of the external potential strength, $U/\mu$, up to the desired
order.

\subsection{Fluctuation Grand Potential}
The GP of fluctuations, described by the quadratic Hamiltonian 
\eqref{H2inhom}, 
can be split into the sum of two terms: 
\begin{equation} 
\Omega_2 = \Omega_2^{(0)} + \delta\Omega_2 .
\end{equation} 
Indeed, the homogeneous contribution $\Omega_2^{(0)} = - \ln \Xi_{0} /\beta$ 
from the partition function $\Xi_0= \tr \{ \exp[-\beta \hat
H_2^{(0)} ] \}$ has been calculated as the second term in eq.~\eqref{eqOmegaHom}
above. Factorizing this known contribution, the complete partition
function belonging to the quadratic Hamiltonian \eqref{H2inhom} can be
written as  
\cite[eq.\ (10.13)]{Bruus2004}
\begin{equation}\label{eqPerturbativeOmega}
\tr \{ \exp[-\beta(\hat H_2^{(0)} + \hat U)]\} = \Xi_0  \mv{\exp[-\textstyle{\int}_0^\beta \rmd \tau \hat U(\tau)]}_0, 
\end{equation}
where the thermal expectation value 
$\langle X \rangle_0 = \tr\{\hat\rho_0 \rmT_\tau X\}$ 
over the Gibbs state 
$\hat\rho_0 = \Xi_0^{-1} \exp\bigl[{-\beta \hat
  H_2^{(0)}}\bigr]$, as well as the Matsubara time evolution
involves only the homogeneous
Hamiltonian. 
Thus, the disorder-produced shift in the GP,  
\begin{equation}
-\beta \delta\Omega_2 = \avg{ \ln  \mv{\exp[-\textstyle{\int}_0^\beta \rmd \tau \hat U(\tau)]}_0},
\end{equation} 
can now be computed straightforwardly by
perturbation theory in powers of 
the effective scattering potential
\eqref{eqInhomHG2}. 
Taking the logarithm leaves us with the connected
correlations, which up to second order read 
\begin{align}
 \beta \, \delta \Omega_2
  =             \Bigl< \int_0^\beta \hspace{-1ex}\rmd\tau \, \avg{\hat U(\tau)} \Bigr>_0
  - \frac{1}{2} \Bigl< \int_0^\beta \hspace{-1ex}\rmd\tau \int_0^\beta \hspace{-1ex}\rmd\tau' \, \avg{\hat U(\tau)\hat U(\tau')} \Bigr>_0^\text{c}.
\end{align}
With the help of Wick's theorem, all correlations can be expressed by 
Matsubara Green functions connecting the matrix elements 
\eqref{eqW} and \eqref{eqY} of $\hat U$. Expanding these in turn  
to second order in the external potential, we find  
\begin{align}\label{eqOmega1V}
\delta \Omega_2^{(2)} &=
          \sum_{\k} \avg{W_{\k \k}^{(2)}} \bigl[\nB{\k} + \tfrac{1}{2}\bigr]\\
        + &\frac{1}{2}\sum_{\k \k'} \left\{
           \avg{\bigl|W_{\k\k'}^{(1)}\bigr|^2} \frac{\nB{\k}-\nB{\k'}}{\ep{\k}-\ep{\k'}}
         - \avg{\bigl|Y_{\k\k'}^{(1)}\bigr|^2} \frac{1+\nB{\k}+\nB{\k'}}{\ep{\k}+\ep{\k'}}
          \right\} .\nonumber
\end{align}
This expression for the disorder-induced correction to the GP of
quantum fluctuations is the central result of this article. Let us
emphasize that our approach, starting from the deformed condensate
background, takes into account \emph{all} contributions that are of
second order in $U$ (and thus goes beyond the method of Huang and Meng
\cite{Huang1992,Kobayashi2002}). 
Furthermore, this result only involves elementary
perturbation theory and does not rely on the replica method. Its
obvious drawback is its perturbative nature. We therefore do not make
any claims concerning the strong-disorder regime, nor do we cover high
temperatures or strongly interacting regimes, where interactions
between excitations become important (see \cite{Lopatin2002,Falco2007}). Here, we rather wish to
provide an account as complete as possible of disorder-effects up to
second order in $U/\mu$ at low temperatures, where Bogoliubov theory
applies.

\subsection{Condensate depletion}
We are now in the position to compute the particle
number shift $\delta N_2=-\partial \delta \Omega_2^{(2)}/\partial \mu$ due
to the disorder. When this number is compared to the number of
particles in the condensate, $\Nc$, we find the additional condensate
depletion caused by the inhomogeneous potential, or 
``potential depletion'' for short. 
In the partial derivative of \eqref{eqOmega1V} with respect to $-\mu$ at fixed $\nc$, we set 
$\mu \approx g \nc \approx g n$ in the end because the expression is already of first order in the dilute-gas parameter and second order in the disorder strength.
We find the  
following collection of identities helpful:
$-\partial \ept{\k}/\partial \mu = 1 $, as well as 
\begin{align}
-\frac{\partial \ep{\k}}{\partial \mu} & = u_\k^2 + v_\k^2,   &   
  - a_\k^{-1} \frac{\partial a_{\k}}{\partial \mu} 
 &= a_\k \frac{\partial a_{\k}^{-1}}{\partial \mu} = \frac{u_\k v_\k}{\ep{\k}}, \displaybreak[0]\\
- \frac{\partial  S_{\k \k'}}{\partial \mu} &= \frac{2 gn_{\k-\k'}}{\mu}, &
- \frac{\partial R_{\k \k'}}{\partial \mu}  &= \frac{2 g{\check n}_{\k-\k'}}{\mu}, \label{SRdmu}
\end{align} 
with $\check n_\q = [\nc^2/\nc(\r)]_\q$.
Applying these to Eqs.\ \eqref{eqW} and \eqref{eqY}, one finds
\begin{subequations} 
\begin{align}
 -\frac{\partial  W_{\k \p} }{\partial \mu}
  &= \frac{a_\k^2 a_{\p}^2{\check n}_{\k-\p} + {n}_{\k-\p}}{2 a_\k a_{\p} \mu/g} 
    + \left[ \frac{u_\k v_\k}{\ep{\k}} + \frac{u_{\p} v_{\p}}{\ep{\p}}\right]Y_{\k \p} ,\\
- \frac{\partial  Y_{\k \p} }{\partial \mu}
  &= \frac{a_\k^2 a_{\p}^2{\check n}_{\k-\p} - {n}_{\k-\p}}{2 a_\k a_{\p} \mu/g} 
    + \left[ \frac{u_\k v_\k}{\ep{\k}} + \frac{u_{\p}
        v_{\p}}{\ep{\p}}\right]W_{\k \p} \label{Ydmu} .
\end{align}
\end{subequations}
Via \eqref{eqW}--\eqref{eqR} and $n_\k = V^{-1}\sum_{\k'} \Phi_{\k-\k'}\Phi_{\k'}$,
we express the right hand sides in terms of the perturbative solution of the Gross-Pitaevskii equation \eqref{smoothing01}--\eqref{smoothing2}.
The relevant expressions up to second order read
\begin{align}
g n_\q^{(1)} &= -2 \mu \Vtil_\q = - g\check n_\q^{(1)}, \label{n1}\\
g n^{(2)}_0  &= \textstyle \sum_\q \epn{\q} |\Vtil_\q|^2, \quad
g\check n^{(2)}_0 = \sum_\q (4\mu-\epn{\q}) |\Vtil_\q|^2, \label{n2}
\end{align}
and
\begin{align}
W^{(1)}_{\k \p} &= \tilde w^{(1)}_{\k \p} \Vtil_{\k-\p}, &
Y^{(1)}_{\k \p} &= \tilde y^{(1)}_{\k \p} \Vtil_{\k-\p}, \\
W_{\k \k}^{(2)} &= \textstyle \sum_{\q} {\bigl|\Vtil_{\q}\bigr|^2} \tilde{w}^{(2)}_{\k, \k+\q}, &
Y_{\k \k}^{(2)} &= \textstyle \sum_{\q} {\bigl|\Vtil_{\q}\bigr|^2} \tilde{y}^{(2)}_{\k, \k+\q},
\end{align} 
with $\tilde w^{(1)}_{\k \p}$ and $\tilde y^{(1)}_{\k \p}$ as given in Ref.\ \cite{Muller2012_momdis} and
\begin{subequations}\label{wy2}
\begin{align}  
  \tilde{w}^{(2)}_{\k, \k+\q}
  &= \left[2\epn{\k} + 3\epn{\q} + (\epn{\k}+\mu) \lambda_{\k \q}
        \right] {\epn{\k}}/{\ep{\k}} ,  \\
  \tilde{y}^{(2)}_{\k, \k+\q}
  &= \Bigl[2\epn{\k} +  \epn{\q} - {\epn{\k}\epn{\q}}/{\mu} + \mu \lambda_{\k \q}\Bigr] {\epn{\k}}/{\ep{\k}}  .
\end{align}
\end{subequations}
The term $\lambda_{\k\q}=(\epn{\k-\q}+\epn{\k+\q}-2\epn{\k}-2\epn{\q})/2\epn{\k}$ vanishes in the present case of a quadratic dispersion relation $\epn{\k}\propto k^2$, but is nonzero in the case of a lattice potential \cite{Gaul2013_bogolattice}.

We can write down the potential depletion as 
\begin{align}\label{eqPotentialDepletion}
 \avg{\delta n^{(2)}} = - \frac{1}{V} \frac{\partial {\delta
     \Omega_2^{(2)}}}{\partial \mu} = \frac{1}{V} \sum_{\q} G(\q)   \avg{{\bigl|U_{\q}\bigr|}^2},
\end{align} 
with a kernel $G(\q) = (2\mu+\epn{\q})^{-2}\sum_\k \tilde M_{\k \k+\q}^{(2)}$ defined in
terms of the envelope 
\begin{align}\label{kernelM} 
 \tilde M^{(2)}_{\k\p} &= v_\k^2 + u_\p^2 
   + (\nB{\k}+\nB{\k}^\prime \tilde w^{(2)}_{\k\p}) (u_\k^2 + v_\k^2) \nonumber\\
   +& \nB{\p} (u_\p^2 + v_\p^2)
   + {u_\k v_\k} (1+2\nB{\k}) \Bigl[ \frac{\tilde y^{(2)}_{\k\p}}{\ep{\k}} - 2 + \frac{\epn{\k-\p}}{\mu}\Bigr] \nonumber\\ 
  -& 2 \frac{1+\nB{\k}+\nB{\p}}{\ep{\k}+\ep{\p}}
    \left[u_\k u_\p+ v_\k v_\p + \frac{u_\k v_\k}{\ep{\k}} \tilde{w}^{(1)}_{\k\p}\right]\tilde{y}^{(1)}_{\k\p}\nonumber\\
  -& \biggl( \nB{\k}^\prime  - \frac{1+\nB{\k}+\nB{\p}}{\ep{\k}+\ep{\p}}  \biggr) (u_\k^2+v_\k^2)  \frac{(\tilde y^{(1)}_{\k\p})^2}{\ep{\k}+\ep{\p}} \nonumber\\ 
  -&2\frac{\nB{\k}-\nB{\p}}{\ep{\k}-\ep{\p}}\left[u_\k u_\p+ v_\k v_\p - \frac{u_\k v_\k}{\ep{\k}}\tilde{y}^{(1)}_{\k\p}\right]\tilde{w}^{(1)}_{\k\p}\nonumber\\
  +&\biggl( \nB{\k}^\prime - \frac{\nB{\k}-\nB{\p}}{\ep{\k}-\ep{\p}} \,
    \biggr) (u_\k^2+v_\k^2) \frac{(\tilde w^{(1)}_{\k\p})^2}{{\ep{\k}-\ep{\p}}} ,
\end{align} 
where $\nB{\k}^\prime=\partial\nB{\k}/\partial\ep{\k}|_{\mu=g\nc}=\beta (\nB{\k}+\nB{\k}^2)$.
Equation \eqref{eqPotentialDepletion} leaves a certain freedom to
exchange $\p$ and $\k$ in the individual components of $\tilde M_{\k \p}^{(2)}$. 
In the spirit of Ref.\ \cite{Muller2012_momdis}, we have used this freedom to write down eq.\ \eqref{kernelM} in a way that allows the identification of  
$\delta n_\k^{(2)} \equiv \sum_\q |\Vtil_\q|^2 \tilde M_{\k,\k+\q}$ with the momentum distribution of the condensate depletion induced by the disorder. 

\begin{figure}[tp]
 {\includegraphics[width=\linewidth]{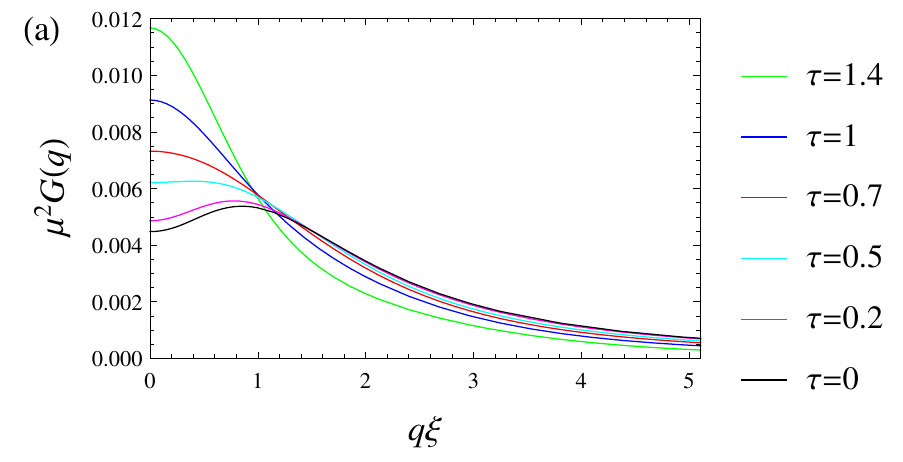}}\\
 {\includegraphics[width=\linewidth]{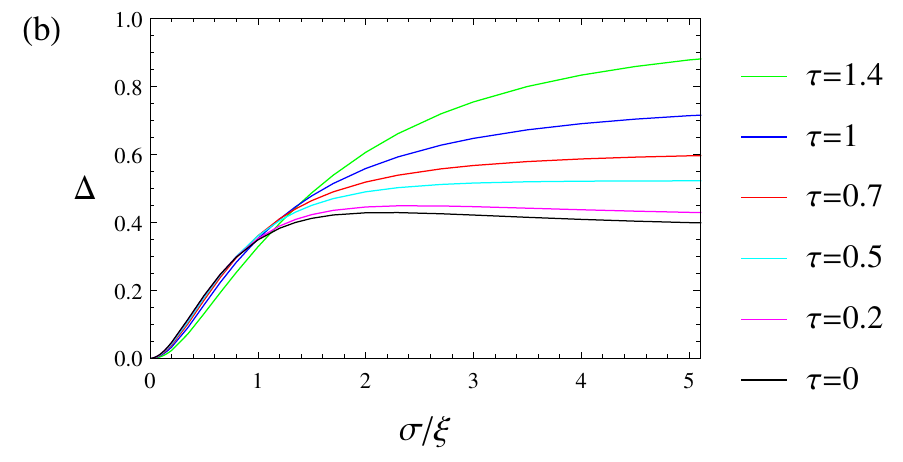}}
\caption{Disorder-induced condensate depletion \eqref{eqPotentialDepletion} for different values of the dimensionless temperature $\tau=\kB T/\mu$. Panel (a) shows the kernel $G(\q)$, whereas panel (b) shows the potential depletion $\avg{\delta N^{(2)}}$ compared to the clean depletion \eqref{deltan0} at zero temperature in units of the square of the dimensionless disorder: 
$\Delta = \avg{\delta n^{(2)}} / [\delta n_0 ({U}/{\mu})^2]$.
\label{figDisorder}}
\end{figure}
Now we evaluate eq.\ \eqref{eqPotentialDepletion} 
in the case of a disorder potential with strength $U$ and correlation length~$\sigma$:
\begin{equation}
 \avg{|U_\q|^2} = V^{-1} U^2 (2\pi)^{3/2} \sigma^3 \exp(-q^2\sigma^2/2) .\label{gaussian}
\end{equation}
Results are shown in Fig.\ \ref{figDisorder}(b) as a function of the disorder correlation length for different temperature values.
In all cases there is an increase of the depletion due to disorder.
At first sight surprisingly, this increase diminishes with temperature in the regime of uncorrelated disorder $\sigma \ll \xi$.
However, the disorder correction as shown in Fig.\ \ref{figDisorder}(b) is expressed in units of the homogeneous quantum depletion at zero temperature \eqref{deltan0}.
Thus, it goes on top of the thermal depletion discussed in Fig.~\ref{figClean}, such that both temperature and disorder deplete the condensate.

We have found [Figs.\ \ref{figClean}(b) and \ref{figDisorder}(b)] that for temperatures up to $\kB T \lesssim \mu$ and for not too strong disorder $U \lesssim \mu$, the depletion $\delta n^{(0)} + \delta n^{(2)}$ remains of the same order of magnitude as the zero-temperature homogeneous depletion $\delta n_0$, which \textit{a posteriori} validates the Bogoliubov method.

Finally, we note that the bulky expression for the envelope function \eqref{kernelM} can be simplified significantly in the Thomas-Fermi regime $\sigma\gg\xi$, where the condensate profile can faithfully follow the variations of the disorder potential on the length scale $\sigma$.
In this case, the disorder correlation \eqref{gaussian} tends to a Dirac $\delta$-function and the potential depletion $\delta n^{(2)}$ is dominated by the diagonal elements $\tilde M_{\k\k}^{(2)}$, which are given as
\begin{align}
 \tilde M_{\k\k}^{(2)} = \frac{(k^2\xi^2-1)(1+2\nB{k} -2 \nB{k}^\prime \ep{k}) + 2 (k^2\xi^2+1) \nB{k}^{\prime\prime} \ep{k}^2}{k\xi(2+k^2\xi^2)^{5/2}}.
\end{align}
Summing over $\k$ then gives $G(0)$ [left edge of Fig.\ \ref{figDisorder}(a)], which is proportional to the depletion in the Thomas-Fermi limit $\sigma/\xi \to \infty$ [right edge of Fig.\ \ref{figDisorder}(b)].

\subsection{Connection to the canonical frame}\label{appCanonical}
At a given value of the chemical potential, the disorder potential draws more particles into the condensate [eq.\ \eqref{gnc_of_muU}].
In Refs.\ \cite{Gaul2011_bogoliubov_long,Muller2012_momdis} the canonical frame is used, where this effect is compensated by a shift $\Delta \mu =-\sum_\q \epn{\q} |\Vtil_\q|^2$ of the chemical potential \cite{Gaul2011_bogoliubov_long},
which results in different second-order expressions \eqref{n2} and \eqref{wy2}. 
In particular, the expression for the potential depletion given in Ref.\ \cite[Eqs.\ (48) and (49)]{Muller2012_momdis} differs slightly from the one given here
in Eqs.\ \eqref{eqPotentialDepletion} and \eqref{kernelM}, even at zero temperature.
Equation \eqref{kernelM} goes over to its canonical form at fixed $\nc$ by replacing $\tilde w^{(2)}_{\k\p}$ and $\tilde y^{(2)}_{\k\p}$ with their canonical expressions (given in Ref.\ \cite{Muller2012_momdis}) and by dropping the term $\epn{\k-\p}/\mu$ in the second line of eq.~\eqref{kernelM}.
In fact, the difference between the two frames can be written as 
\begin{equation}
\delta n^{(2)}_{\rm canonical} - \delta n^{(2)}_{\rm gc} = \Delta \mu \frac{\partial \delta n^{(0)}}{\partial \mu}, 
\end{equation}
where $\delta n^{(0)} = V^{-1} \sum_\k [v_\k^2 + \nB{\k}(u_\k^2+v_\k^2)]$ is the homogeneous depletion at finite temperature.
With this shift, the kernel functions shown in Fig.\ \ref{figDisorder}(a) would begin with the same values at $q=0$ and then decrease monotonically as function of $q$ without crossing each other. Likewise, the disorder-induced depletion of Fig.\ \ref{figDisorder}(b) would become monotonic without crossings; the zero temperature curve takes the same form as in Figure 4 of \cite{Muller2012_momdis}.

\section{Conclusions}
We have applied the grand-canonical formulation to the problem of disordered Bose-Einstein condensates,
which brings conceptual advantages over the conventional canonical frame.
Once the grand potential is determined, one obtains relevant physical quantities by differentiation.
The condensate mode $\Phi(\r)$ plays a special role in the grand-canonical Bogoliubov approach.
In principle, it has to minimize the grand potential, which includes a back action of the excitations on the condensate.
For the main work of this article, we have chosen the equivalent `semicanonical' approach, where one keeps the condensate as a parameter that is inserted only at the end of the calculation (i.e., after taking derivatives). To the desired precision, it is then sufficient to determine the condensate mode by minimizing the ground-state energy.

Concerning physical results, we have mainly focused on the speed of sound, the compressibility, as well as on the particle fractions \emph{condensate fraction} and \emph{condensate depletion}.
In particular, we have reproduced previous results \cite{Muller2012_momdis} on the disorder-induced condensate depletion from the perspective of the grand-canonical picture and have extended them to the case of finite temperatures.

\appendix
\section{Grand-canonical condensate density with beyond-mean--field
  corrections} 
\label{appHom}

In Sec.~\ref{secHom}, we have determined the condensate density $\nc$
by minimizing the ground-state energy $E_0$ at fixed $\mu$. This
amounts to determining the condensate density, once and for all, at zero
temperature. When the temperature is raised, then of course thermal
excitations will appear, which deplete the condensate. This
effect can be explicitly accounted for by determining $\nc$ directly from the
GP $\Omega$ at finite temperature.  Then, using this (now $\mu$ and
$T$ dependent) solution, one has a GP $\Omega(\mu,T)$ that depends
only on $\mu$, and not separately on $g\nc$ anymore. The total density
then derives by differentiation with respect to this $\mu$
alone. This proper grand-canonical procedure yields the same results than the `semicanonical'
method used in Sec.~\ref{secHom} above, as demonstrated in the
following.  

Requiring that the homogeneous GP, eq.~\eqref{eqOmegaHom},  be stationary, $\partial
\Omega/\partial\nc|_\mu=0$, yields the condensate density  
\begin{equation} \label{ncalt}
\nc = \frac{\mu}{g} - \frac{5\sqrt{2}}{12\pi^2} \frac{1}{\xi^3}- \int  \frac{{\rm d}^3 k}{(2\pi)^3}  \nB{\k} \left.\frac{\partial \ep{\k}}{\partial
 g \nc}\right|_\mu
\end{equation}   
with now a $T$-dependent contribution. 
Inserting this solution into \eqref{E0} yields actually \emph{the
  same} ground-state energy \eqref{E0ofmu} as before. The reason is
that the beyond-mean-field correction $\nc = (\mu/g) + \Delta \nc$ does not
contribute there to lowest order, since this correction is only used
in the mean-field term $H_0$, eq.~\eqref{Ec}, for which   
\begin{equation}
(g \nc -2\mu)g \nc = - (\mu - g\Delta \nc)(\mu + g\Delta\nc) = -\mu^2   
\end{equation} 
to the order considered. Thus, the ground-state energy $E_0(\mu)$ is
unchanged, just as the GP, eq.~\eqref{eqOmegaHom}. The difference now
is that the excitation energy inside the fluctuations is to be taken at
the Bogoliubov dispersion $\epB{\k}$ from eq.~\eqref{bogospec}
as function of $\mu$ alone. Thus, the total number of particles (for
the same $\mu$) is now different, namely, 
\begin{equation} \label{n_hom_ofmu_alternative}
n = 
-\frac{1}{V}\frac{\partial E_0}{\partial\mu} -  \int
\frac{{\rm d}^3 k}{(2\pi)^3}  \nB{\k} \frac{\partial \epB{\k}}{\partial
  \mu},   
\end{equation}
 with $\partial \epB{\k}/\partial\mu =
 \epn{\k}/\epB{\k}$. However, also the condensate density \eqref{ncalt}  is
 now temperature dependent. The difference between these two
 densities is the depleted density 
\begin{equation}
\delta n = n-\nc = \delta n_0 -\int \frac{{\rm d}^3 k}{(2\pi)^3}
\nB{\k}\left(\frac{\partial \epB{\k}}{\partial
  \mu} -\left. \frac{\partial \ep{\k}}{\partial
 g \nc}\right|_\mu
\right) 
\end{equation} 
with the zero-temperature depletion $\delta n_0$ given by eq.~\eqref{deltan0}. As
for the thermal depletion, the dispersion relation is such that the difference of derivatives
appearing there is precisely the result we had before:  
\begin{equation}
 \frac{\partial \epB{\k}}{\partial
  \mu} - \left.\frac{\partial \ep{\k}}{\partial
 g \nc}\right|_\mu = \left.\frac{\partial \ep{\k}}{\partial
 \mu}\right|_{g\nc}. 
\end{equation} 
Thus, \eqref{eqParticleNumberTh} still holds as before, and all
zero-$T$ results are identical anyway. 
It is largely a matter 
of taste whether one wants to have a $T$-dependent contribution to
$\nc$ or not, and whether one wants to have the GP depend really on
$\mu$ alone. Both approaches, strict grand canonical and
`semicanonical', are equivalent.

\nocite{apsrev41Control}

\bibliography{references_16}

\end{document}